\documentstyle[12pt,epsfig]{article}

\newcommand{\be}{\begin{equation}}
\newcommand{\ee}{\end{equation}}
\begin{document}  
\topmargin 0pt
\oddsidemargin=-0.4truecm
\evensidemargin=-0.4truecm
\renewcommand{\thefootnote}{\fnsymbol{footnote}}
\newpage
\setcounter{page}{0}
\begin{titlepage}   
\vspace*{-2.0cm}  
\begin{flushright}
FT-UM-TH-03-10 \\
CERN-TH-03-134 \\
\end{flushright}
\vspace*{0.1cm}
\begin{center}
{\Large \bf KamLAND and Solar Antineutrino Spectrum \footnote{Based on talk given at NANP'03, JINR Dubna, Russia, June 2003.}} \\ 

\vspace{1cm}

{\large 
Bhag C. Chauhan$^a$,
Jo\~{a}o Pulido$^a$,
 and E.~Torrente-Lujan$^b$\\
\vspace{0.15cm}
{ $^a$ {\small \sl Centro de F\'\i sica das Interac\c c\~oes Fundamentais (CFIF) \\
Instituto Superior T\'ecnico,
Av. Rovisco Pais, P-1049-001 Lisboa, Portugal.}\\
$^b$ {\small\sl Departamento de Fisica, 
Universidad de Murcia, Murcia, Spain.}\\
$^{b}$ {\small\sl CERN-TH, CH-1211 Geneve 23, Switzerland.}
}}
\end{center}
\vglue 0.6truecm

\begin{abstract}
We use the recent KamLAND observations to predict the solar antineutrino 
spectrum at some confidence limits. 
We find that a scaling of the antineutrino probability with respect
to the magnetic field profile --in the sense that the same probability
function can be reproduced by any profile with a suitable peak field value--
can be utilised to obtain a general shape of the solar antineutrino spectrum. 
This scaling and the upper bound on the solar antineutrino
event rate, that can be derived from the data, lead to: 1) an upper bound 
on the solar antineutrino flux,  2) the prediction of their energy spectrum, 
as the normalisation of the spectrum can be obtained from the total number of 
antineutrino events recorded in the experiment.    
We get $\phi_{\bar\nu}<3.8\times 10^{-3}\phi(^8B)$ or $\phi_{\bar\nu}
<5.5\times 10^{-3}\phi(^8B)$ at 95\% CL, assuming Gaussian or Poissonian 
statistics, respectively. 
And for 90\% CL these become $\phi_{\bar\nu}<3.4
\times 10^{-3}\phi(^8B)$ and $\phi_{\bar\nu}<4.9\times 10^{-3}\phi(^8B)$.
It shows an improvement by a factor of 3-5 with respect to existing bounds.
These limits are quite general and independent of the detailed structure of the
magnetic field in the solar interior. 
\end{abstract}

\end{titlepage}   
\renewcommand{\thefootnote}{\arabic{footnote}}
\setcounter{footnote}{0}

\section{Introduction and Motivation}
Assuming CPT invariance, the long-standing solar neutrino problem (SNP) seems to be resolved after the recent results from the KamLAND experiment \cite{Eguchi:2002dm}. The results confirm that the most favoured large mixing angle (LMA) solution is chosen by nature at the dominant level. Just before KamLAND data, there were also some other possible solutions equally favoured by the solar neutrino data, namely LOW, VO \cite{Aliani:2002ma,klothers} and Spin Flavour Precession (SFP) \cite{SV,LM,Ak} etc. It had been obvious that this deficit had to rely on 'non-standard' neutrino properties, namely neutrino mass and magnetic moment. SFP is based on the interaction of the neutrino magnetic moment with the solar magnetic field, was giving as good data fits as that of the most favourable LMA oscillations \cite{generalrandom}.    

If neutrinos are endowed with a sizeable transition magnetic moment, the SFP, although certainly not playing the major role in the solar neutrino deficit, may still be present as a subdominant process. Its signature will be the appearance of solar antineutrinos \cite{LM,Ak1,alianiantinu} which result from the combined effect of the vacuum mixing angle $\theta$ and the transition magnetic moment $\mu_{\nu}$ converting neutrinos into antineutrinos of a different flavour. 

The subdominant effect of SFP has been studied recently 
by \cite{Akhmedov:2002mf}.
The question of what can be learned about the strength and coordinate
dependence of the solar magnetic field in relation to the current upper 
limits on the solar $\bar\nu_{e}$ flux was addressed. The system of equations
describing neutrino evolution in the sun was solved analytically in 
perturbation theory for small $\mu_{\nu}{B}$, the product of the neutrino 
magnetic moment by the solar field. The three oscillation scenarios with the 
best fits were considered, namely LMA, LOW and vacuum solutions. In particular,
 for LMA it was found that the antineutrino probability depends only on the 
magnitude of the magnetic field in the neutrino production zone. 

In the combined effect there are essentailly two ways to produce antineutrinos.
This can be schematically shown as 

\be
{\nu_e}_L \rightarrow {\nu_{\mu_L}} \rightarrow {\bar\nu_{e_R}},
\ee
\be
{\nu_e}_L \rightarrow {\bar\nu_{\mu_R}} \rightarrow {\bar\nu_{e_R}}~~.
\ee

In (1) the oscillations acting first and SFP in the next step; however, in (2) neutrinos first undergo SFP and then oscillate into antineutrinos. Oscillations and SFP can either take place in the same spatial region, or be spatially separated. 

As the antineutrino production inside the sun is strongly suppressed \cite{Akhmedov:2002mf} they can be produced on the way of neutrino propagation to earth. 
The processes corresponding to eq. (1), can be disregarded as the magnetic field in the region between the sun and the earth is negligibly small. For the processes via the squence (2), the probability that a ${\nu_e}_L$ 
produced inside the sun will reach the earth as a $\bar\nu_{e_R}$ can be written as
\be
P(\nu_{e_L} \rightarrow \bar\nu_{e_R})=P(\nu_{e_L} \rightarrow 
\bar\nu_{\mu_R};R_S) \times 
P(\bar\nu_{\mu_R} \rightarrow \bar\nu_{e_R};R_{es})~~,
\ee
in which the first term is the SFP probability, $R_S$ is the solar radius
and the second term is given by the well known formula for vacuum oscillations
\be
P(\bar\nu_{\mu_R} \rightarrow \bar\nu_{e_R};R_{es})={\sin}^{2}2\theta 
~~{\sin}^{2}\! \left(\frac{\Delta m^{2}}{4E}R_{es}\right)~~,
\ee
which can be approximated by '1/2' for the LMA parameters, 
$\Delta m^{2}=6.9\times10^{-5}eV^{2}$, $\sin^2 2\theta=1$ \cite{Eguchi:2002dm},  taking the $\bar\nu_{\mu_R} \rightarrow \bar\nu_{e_R}$ vacuum oscillations 
to be in the averaging regime. Here $R_{es}$ is the distance between the sun and the earth and the rest of the notation is standard. 

The SFP amplitude in perturbation theory for small $\mu B$ 
is obtained by \cite{Akhmedov:2002mf} as 
\be
A(\nu_{e_L} \rightarrow \bar\nu_{\mu_R})= \frac{\mu B(r_i) \sin^{2}\theta (r_i)}{g^{'}_2(r_i)}~~,
\ee

$g^{'}_2$ is a function of $V_e,~V_{\mu},~\theta,~\Delta m^2/4E.$

In their work \cite{Akhmedov:2002mf}, they assumed all the neutrinos are being produced at the same point ($x=0.05R_S$), where $^8 B$ neutrino production is peaked. However, we have considered a more realistic case recently \cite{bje}, in which we convolute the neutrino production distribution spectrum with the solar magnetic field profile. In this way, the overall antineutrino production probability is obtained as 
\be
P(\nu_{e_L} \rightarrow \bar\nu_{e_R})= \frac{1}{2} \int |A(\nu_{e_L} 
\rightarrow \bar\nu_{\mu_R})|^2 f_{B}(r_i)dr_i~~,
\label{prob}
\ee

where $f_{B}$ represents the neutrino production distribution function for Boron neutrinos \cite{Bahc} and the integral extends over the whole production region. 

Owing to this integration, the energy shape of probability (\ref{prob}) is largely insensitive to the magnetic field profile function. As a result the infinite ambiguities about the shape of solar field profile are omitted.

Now, we can define a parameter $k$ such that

\be
P_{\overline{\nu}}[{B(r)}]~=~ k P_{\overline{\nu}}^0~~.
\ee

This scaling factor $k$ relates one probability function to the other, via:
\be
k=\frac{\int \left(\frac{B(r_i)\sin^2\theta(r_i)}{g^{'}_{2}(r_i)}\right)^2 f_B (r_i)dr_i}
{\int \left(\frac{B^{0}(r_i)\sin^2\theta(r_i)}{g^{'}_{2}(r_i)}\right)^2
 f_B (r_i)dr_i}~~.
\ee
 
The effect of parameter $k$ can be seen in the upper panel of 
fig.2 in \cite{bje}. One can obtain the same probability function for a 
suitable choice of peak field value. 
In this case the probability curves differ only slightly in their 
shapes while their normalizations are the same. 
 
Using the fact of profile independence in the shape of antineutrino production 
probability, it is possible to extract a general 
solar antineutrino spectrum from the recent KamLAND observations at some confidence levels. However, the true normalization of spectrum can be obtained from the total solar antineutrino events seen by KamLAND in future.

\section{Solar Antineutrinos in KamLAND}
KamLAND is sensitive to antineutrinos of $E_{\overline{\nu}}>1.8~MeV$ via 
a reaction of positron production
\be
\overline{\nu_e}+p~\rightarrow~n+ e^{+}~~.
\ee

This experiment detects antineutrinos from several reactors around the 
Kamioka mine in Japan. In fact, it is also sensitive to the antineutrinos 
coming from the sun.
However, the recent KamLAND data doesn't indicate any solar antineutrino 
signal, yet it couldn't rule out the presence of solar antineutrinos in 
the huge background of neutrinos from reactors and other sources.

The positron event rate in the KamLAND experiment originated from 
solar antineutrinos can be written as
\be
S=Q_0\int_{E_e^0}^{\infty} dE_e \int_{E_m}^{E_M} \epsilon (E^{'}_e) R(E_e,E^{'}_e) 
\phi_{\bar\nu}(E) \sigma(E)dE.
\label{erate}
\ee
Where $Q_0$ is a normalization factor which takes into account the 
number of atoms of the 
detector and its live time exposure \cite{Eguchi:2002dm} and 
$E$ is the antineutrino energy, related to the physical positron energy 
by $E^{'}_e=E-(m_N-m_P)$ 
to zero order in $1/M$, the nucleon mass. We thus
have $E_m=1.804MeV$, while the KamLAND puts a lower energy cut at $E_e^0=2.6MeV$ in order to eliminate the naughty background of geo-neutrinos. The functions
$\epsilon$ and $R$ denote the detector efficiency and $R(E_e,E^{'}_e)$ is the 
Gaussian energy resolution function of the detector.

We assume a  408 ton fiducial mass and  the detection efficiency is 
taken independent of the energy \cite{Eguchi:2002dm},
 $\epsilon\simeq 80\%$, which amounts to $162$ ton.yr of antineutrino 
data. The antineutrino cross section $\sigma(E)$ was taken from 
ref.\cite{Vogel:1999zy}
and we considered energy bins of size $E_e=0.425\ MeV$ 
in the KamLAND observation range $(2.6-8.125)~MeV$ \cite{Eguchi:2002dm}.

Taking in to account the previous expression for $k$ and the near invariance of the probability shape one can interpret the expected solar antineutrino signal in KamLAND as \cite{bje}

\be
S_{\overline{\nu}}[{B(r)}]~=~ k S_{\overline{\nu}}^0~~,
\ee

 where $S_{\overline{\nu}}^0$ is the antineutrino signal for a reference peak value for a reference field profile. We make use of this behavior to obtain upper limits on the total antineutrino flux and the $\mu_{\nu} B_{0}$.

We apply Gaussian probabilistic considerations to the global rate in the whole energy range, $E_{\nu}=(2.6-8.125)~MeV$, and Poissonian considerations to the event content in the highest energy bins ($E_e> 6 $ MeV) where KamLAND observes no signal. 

\be
S^{sun}_{\overline{\nu}}~=~ S_{obs}-S_{react}(LMA). 
\ee

Where $S_{obs}=54.3\pm 7.5 $ and $S_{react}(LMA)$ is the signal expected for the best fit parameters of KamLAND ($6.9\times10^{-5}eV^{2},1)$~ $=~ 49\pm 1.3 $. 

We obtain  at 90 (95)\% CLs' 

\be
S_{obs}-S_{react}=k S^{0}_{\bar\nu} < 17.8~ (20.2)~~. 
\ee

As a result we obtain an upper bounds on $\mu_{\nu} B_{0}$ and on the solar antineutrino flux at $90\% ~CL$:

\be
\mu_{\nu} B_{0}~<~5.16 \times 10^{-19}~~MeV;~~
\phi_{\bar\nu}~<~0.0034~~\phi(^8B)
\ee

and at $95\% ~CL$:
\be
\mu_{\nu} B_{0}~<~5.5 \times 10^{-19}~~MeV;~~
\phi_{\bar\nu}~<~0.0038~~\phi(^8B)
\ee

We can similarly and independently apply Poisson statistics to the five highest energy bins ($E>6~MeV$) of the KamLAND experiment where the expected signal from oscillating neutrinos with LMA parameters is negligibly small. 

If no events are observed and in particular no background is observed, the unified intervals \cite {Hagiwara:fs} $[0,\epsilon_{CL}]$ are $[0,2.44]$ at $90\% ~CL$ and $[0,3.09]$ at $ 95\% ~CL$.

So the bounds at $90 \% ~CL$:

\be
\mu_{\nu} B_{0}~<~2.51 \times 10^{-19}~~MeV;~~
\phi_{\bar\nu}~<~0.0049~~\phi(^8B)
\ee

and at $95 \% ~CL$:

\be
\mu_{\nu} B_{0}~<~2.82 \times 10^{-19}~~MeV;~~
\phi_{\bar\nu}~<~0.0055~~\phi(^8B)
\ee

Bounds on $\phi_{\bar\nu}$ at $90 \% ~CL$ show an improvement by a factor of 3-5 with respect to existing bounds from SK \cite{sk}.

\section{Solar Antineutrino Spectrum}
As we have seen that a sufficient magnitude of solar magnetic field makes Spin Flavour Precesion responsible for the production of solar antineutrinos, in the combined action with neutrino oscillations. If SFP is happening inside the sun, as a subdominant effect, along with the LMA oscillations, then inspite of a little knowledge of the solar magnetic fields, it has been found possible to extract the shape of the energy spectrum of the solar antineutrinos produced.  

We use the fact that, for different field profiles the probability curves will
differ only slightly in their shape if they lead to the same number of events.
So for a given number of events the probability curves are essentially the same, regardless of the field profile. As a result the similar, solar field independent, effect can be seen for the spectral flux. 
In other words, from the near independence of antineutrino production probability on the magnetic field profile, there results the near independence of the antineutrino spectrum $\phi_{\overline{\nu}}(E)$ on the profile as any profile can produce the same spectral flux for suitable values of the  peak fields.

Solar antineutrino spectral flux $\phi_{\overline{\nu}}(E)$ can be written as a product $\phi_{\overline{\nu}}^{0} \times f(E)$, where $\phi_{\overline{\nu}}^{0}$ is the total flux and $f(E)$ is some function of energy normalized to one.
On the other hand $\phi_{\overline{\nu}}(E)=\phi_{B}(E) \times P(E)$, where  $\phi_{B}(E)$ is the Boron neutrino flux and $P(E)$ is the antineutrino appearance probability. We know the $\phi_{B}(E)$ from \cite{Bahc} and the $P(E)$ from \cite{bje} and using these functions we can obtain the profile independent and a genaral spectral flux, $\phi_{\overline{\nu}}(E)$, for solar antineutrinos. 

\begin{figure}[h]
\setlength{\unitlength}{1cm}
\begin{center}
\hspace*{-1.6cm}
\epsfig{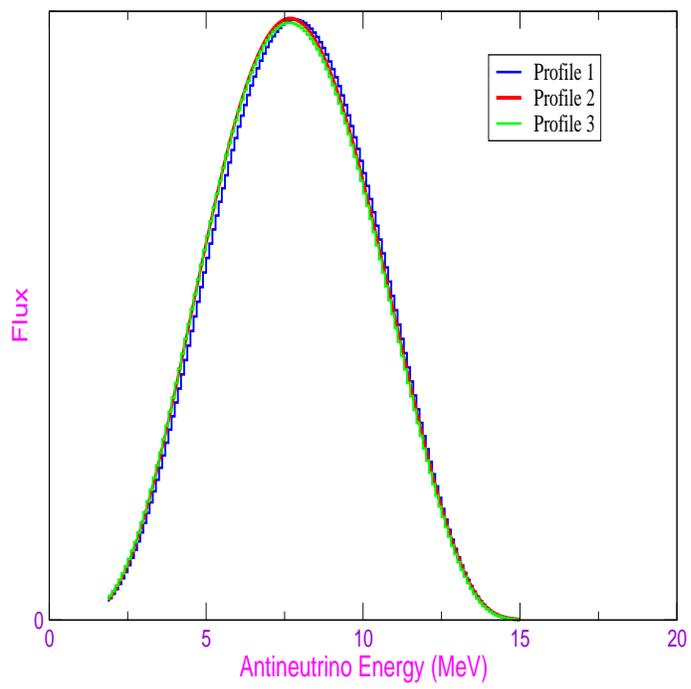}
\end{center}
\caption{ \it Solar antineutrino spectrums for three quite different solar field profiles.} 
\label{fig1}
\end{figure}

In the figure \ref{fig1}, three curves of solar antineutrino spectral flux corresponding to the three quite different solar field profiles \cite{bje}, have been plotted for an arbitrary normalisation of flux. The generality and the profile independence of the spectral flux can be clrearly seen in the figure, as they all are nearly super-imposing of each other and  the energy spectrum of the expected solar antineutrino flux will be nearly the same for any kind of profile. 

\begin{figure}[h]
\setlength{\unitlength}{1cm}
\begin{center}
\hspace*{-1.6cm}
\epsfig{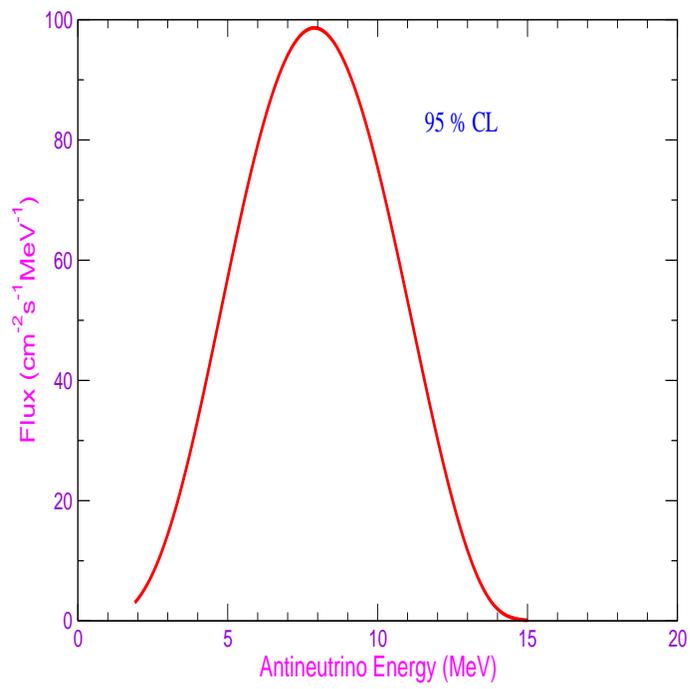}
\end{center}
\caption{ \it Solar antineutrino spectrum normalised to 95\% CL.} 
\label{fig2}
\end{figure}

We have used the confidence limits on the solar antineutrino events obtained from the recent KamLAND observations of first 145 days of data taking \cite{Eguchi:2002dm}, in order to normalize the antineutrino spectrum. In figure \ref{fig2}, the solar antineutrino spectrum is shown for the events normalised to 95\% CL using more democratic case of Poissonian statistics.

\begin{figure}[h]
\setlength{\unitlength}{1cm}
\begin{center}
\hspace*{-1.6cm}
\epsfig{file=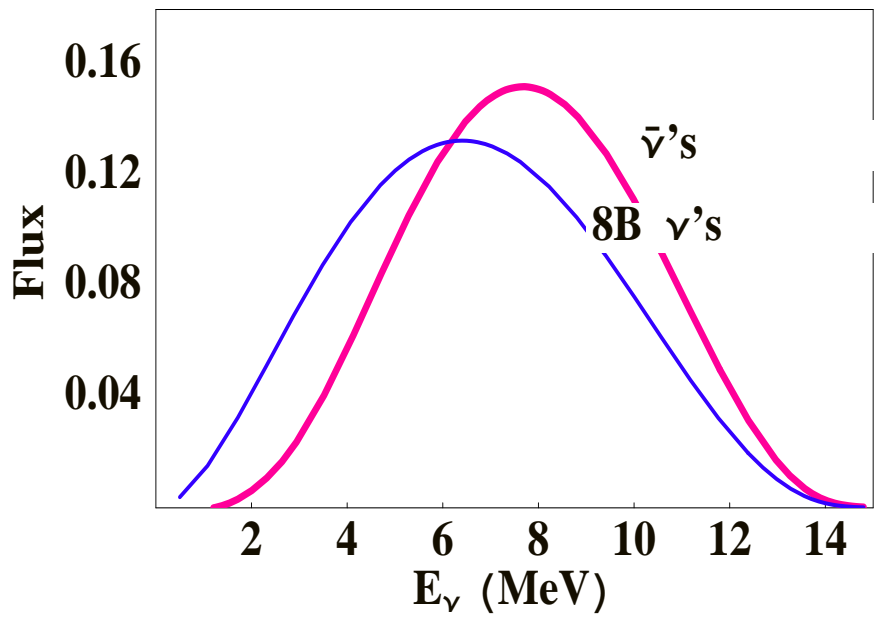,height=9.0cm,width=12.0cm,angle=0}
\end{center}
\caption{\it The expected solar antineutrino spectrum and the 
$^8 B$ neutrino one \cite{Bahc}, both normalized to unity, showing the 
peak shift and the distortion introduced by the antineutrino probability.}
\label{fig3}
\end{figure}

The shape of the spectrum, thus obtained, is not exactly same to that of the parent $^8B$-neutrinos \cite{Bahc} but with a perturbation introduced by the antineutrino production probability function, $P(E)$ \cite{bje}. Solar antineutrino spectrum and $^8B$-neutrino one, both normalized to unity are shown for comparison in figure \ref{fig3}.

\section{Conclusions}
After the recent KamLAND results the magnetic moment solution, SFP, is ruled out as a dominant effect for the solar neutrino problem. It would be interesting to investigate the subdominant effect of SFP as a possible
signature of an observable $\bar\nu_e$ flux in the solar neutrino signal.
Observation of solar antineutrinos in KamLAND will tell us that SFP is occuring in sun.

We used an important result of our recent paper \cite{bje} --antineutrino production probability function is nearly independent of the unknown solar field profile-- and extract a general energy spectrum for solar antineutrinos. In other words, a field profile independent spectrum for solar antineutrinos has been derived. However its normalization can be only achieved through the total observed flux in the experiment. The shape of the solar antineutrino spectrum shows a peak shift and distortion relative to the $^8B$-neutrino spectrum.   

We also obtain the upper bounds on the solar antineutrino flux 
and $\mu_{\nu} B$.
Assuming  Gaussian and Poissonian statistics;  
$\phi_{\bar\nu}<3.8\times 10^{-3}\phi(^8B)$ and $\phi_{\bar\nu}<5.5\times 
10^{-3}\phi(^8B)$ at 95\% CL, respectively. 
For 90\% CL, we found $\phi_{\bar\nu}<3.4\times 10^{-3}\phi(^8B)$
and $\phi_{\bar\nu}<4.9\times 10^{-3}\phi(^8B)$, which shows an improvement 
relative to previously existing bounds from SuperKamiokade \cite{sk} 
by a factor of 3-5. The upper bounds on 
$\mu_{\nu} B_{0}~<~2.51 \times 10^{-19}~~MeV$ at $90\%~CL$ and $\mu_{\nu} B_{0}~<~2.82 \times 10^{-19}~~MeV$ at $95\%~CL$ has been also deduced.
All these bounds are independent of the detailed magnetic field 
profile in the core and radiative zone.

\section*{Acknowledgements}
 The work of BCC  was supported by Funda\c{c}\~{a}o para a Ci\^{e}ncia e a 
Tecnologia through the grant SFRH/BPD/5719/2001. E.T-L  acknowledges many useful conversations with P. Aliani, M. Picariello and V. Antonelli, the hospitality of the CFIF (Lisboa) and the  financial  support of the  Spanish CYCIT  funding agency.

\vspace{1cm}  

\end{document}